\documentclass[showpacs,amsmath,amssymb,prl,twocolumn,prl]{revtex4-1}
\usepackage{graphicx}
\usepackage{soul}
\usepackage[colorlinks=true,citecolor=blue,linkcolor=magenta]{hyperref}
\usepackage[usenames]{color}
\usepackage[cspex,bbgreekl]{mathbbol}
\usepackage{relsize}

\newcommand{\ket}[1]{\left| #1 \right>} 
\newcommand{\si}{Supplementary Information}

\newcommand {\grsim} {\ {\raise-.5ex\hbox{$\buildrel>\over\sim$}}\ }
\newcommand {\lessim} {\ {\raise-.5ex\hbox{$\buildrel<\over\sim$}}\ } 
\newcommand {\ii} {i}

\begin{document}

\title{Realization of the Hofstadter Hamiltonian with ultracold atoms in optical lattices}

\author{M.~Aidelsburger$^{1,2}$, M.~Atala$^{1,2}$, M.~Lohse$^{1,2}$, J.~T.~Barreiro$^{1,2}$, B.~Paredes$^{3}$ and I.~Bloch$^{1,2}$}

\affiliation{$^{1}$\,Fakult\"at f\"ur Physik, Ludwig-Maximilians-Universit\"at, Schellingstrasse 4, 80799 M\"unchen, Germany\\
$^{2}$\,Max-Planck-Institut f\"ur Quantenoptik, Hans-Kopfermann-Strasse 1, 85748 Garching, Germany\\
$^{3}$\,Instituto de F\'{i}sica Te\'{o}rica CSIC/UAM \\C/Nicol\'{a}s Cabrera, 13-15
Cantoblanco, 28049 Madrid, Spain}


\begin{abstract} 
We demonstrate the experimental implementation of an optical lattice that allows for the generation of large homogeneous and tunable artificial magnetic fields with ultracold atoms. Using laser-assisted tunneling in a tilted optical potential we engineer spatially dependent complex tunneling amplitudes. Thereby atoms hopping in the lattice accumulate a phase shift equivalent to the Aharonov-Bohm phase of charged particles in a magnetic field. We determine the local distribution of fluxes through the observation of cyclotron orbits of the atoms on lattice plaquettes, showing that the system is described by the Hofstadter model. Furthermore, we show that for two atomic spin states with opposite magnetic moments, our system naturally realizes the time-reversal symmetric Hamiltonian underlying the quantum spin Hall effect, i.e. two different spin components experience opposite directions of the magnetic field. \end{abstract}

\maketitle

Ultracold atoms in optical lattices constitute a unique experimental setting to study condensed matter Hamiltonians in a clean and well controlled environment \cite{bloch2008many}, even in regimes not accessible to typical condensed matter systems~\cite{reviewsbeyond}. Especially intriguing is their promising potential to realize and probe topological phases of matter, for example, by utilizing the newly developed quantum optical high-resolution detection and manipulation techniques~\cite{qs,singleatoms}. One compelling possibility in this direction is the quantum simulation of electrons moving in a periodic potential exposed to a large magnetic field, described by the Hofstadter-Harper Hamiltonian~\cite{harper1955single,hofstadter1976energy}. For a filled band of fermions, this model realizes the paradigmatic example of a topological insulator that breaks time-reversal symmetry -- the quantum Hall insulator. Moreover, the atomic realization of time-reversal symmetric topological insulators based on the quantum spin Hall effect~\cite{bernevig2006spin} promises new insights for spintronic applications.

The direct quantum simulation of orbital magnetism in ultracold quantum gases is, however, hindered by the charge neutrality of atoms, which prevents them from experiencing a Lorentz force. Overcoming this limitation through the engineering of synthetic gauge potentials is currently a major topic in cold-atom research. Artificial magnetic fields were first accomplished using the Coriolis force in a rotating atomic gas~\cite{madison2000vortex,schweikhard2004rapidly} and later by inducing Berry's phases through the application of Raman lasers~\cite{lin2009synthetic,dalibard2010artificial}. Recently, staggered magnetic fields in optical lattices were achieved using laser induced tunneling in superlattice potentials~\cite{aidelsburger2011experimental} or through dynamical shaking~\cite{Struck21072011}. In one dimension, tunable gauge fields have been implemented in an effective ``Zeeman lattice'' ~\cite{jimenezgarcia2012tunable} and using periodic driving~\cite{struck2012tunable}. Furthermore, the free-space spin Hall effect was observed using Raman dressing~\cite{Beeler2012spinhall}. Despite intense research efforts, 2D optical lattices featuring topological many-body phases have so far been beyond the reach of experiments.

\begin{figure}[t!]
\includegraphics[width=\linewidth]{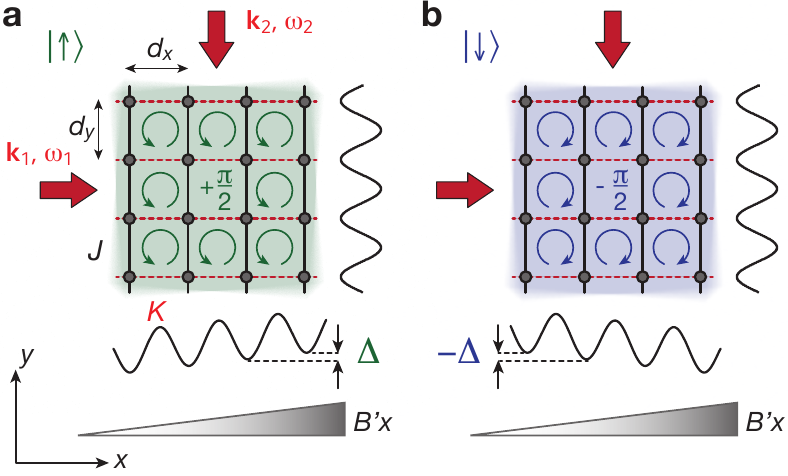}
\vspace{-0.cm} \caption{Experimental setup. The experiment consists of a 3D optical lattice, where the vertical lattice isolates different planes. The lattice constants within each plane are $|\mathbf{d}_i|=\lambda_i/2$, with $i={x,y}$. Along $y$, bare tunneling occurs with strength $J$, while tunneling along $x$ is inhibited by a magnetic field gradient $B'$, which introduces an energy offset between neighboring sites of \textbf{(a)} $\Delta$ for $\ket{\uparrow}$ atoms and \textbf{(b)} $-\Delta$ for $\ket{\downarrow}$ atoms. An additional pair of laser beams (red arrows) with wave vectors $|\mathbf{k}_{1}|\simeq|\mathbf{k}_{2}|=2\pi/ \lambda_K$ and frequency difference $\omega=\omega_1-\omega_2=\Delta/\hbar$ is used to restore resonant tunneling with complex amplitude $K$. This realizes an effective flux of \textbf{(a)} $\Phi=\pi /2$ for $\ket{\uparrow}$ atoms and \textbf{(b)} $-\Phi$ for $\ket{\downarrow}$ atoms. \label{Fig_Scheme}}
\end{figure}

In this Letter, we demonstrate the first experimental realization of an optical lattice that allows for the generation of large tunable homogeneous artificial magnetic fields. The technique is based on our previous work on staggered magnetic fields~\cite{aidelsburger2011experimental}. The main idea is closely related to early proposals by Jaksch and Zoller~\cite{jaksch2003creation} and subsequent work~\cite{gerbier2010gauge,mueller2004artificial}. However, it does not rely on the internal structure of the atom, which makes it applicable to a larger variety of atomic species including fermionic atoms like $^6$Li and $^{40}$K. We use laser-assisted tunneling in a tilted optical lattice through periodic driving with a pair of far-detuned running-wave beams~\cite{kolovsky2011creating,bermudez2011gauge}. In contrast to techniques based on near-resonant laser beams, heating of the atomic cloud due to spontaneous emission is negligible \cite{supplements}. The position-dependence of the on-site modulation introduced by the running-wave beams leads to a spatially-dependent complex tunneling amplitude. Therefore, an atom hopping around a closed loop acquires a non-trivial phase, which mimics an Aharonov-Bohm phase. In our setup we realize a uniform effective flux of $\Phi=\pi/2$ per plaquette, whose value is fully tunable. We study resonant laser-assisted tunneling in the tilted optical potential and reveal the local distribution of fluxes by partitioning the lattice into isolated four-site square plaquettes. Furthermore, we show that  for two spin-states with opposite magnetic moments, $\ket{\uparrow}$ and $\ket{\downarrow}$, our coupling scheme directly gives rise to a non-Abelian SU(2) gauge field that results in opposite magnetic fields for $\ket{\uparrow}$ and $\ket{\downarrow}$ particles. In the presence of such a gauge field the tight-binding Hamiltonian is time-reversal symmetric and corresponds precisely to the one underlying the quantum spin Hall effect~\cite{bernevig2006spin,goldman2010realistic}.

Our experimental setup consists of an ultracold gas of $^{87}$Rb atoms held in a three-dimensional optical lattice created by three mutually orthogonal standing waves of laser light at wavelengths $\lambda_x=\lambda_y=767\,\textrm{nm}$ and $\lambda_z=844\,\textrm{nm}$. The depth of the lattice along $z$ is chosen deep enough to suppress tunneling between individual planes, typically $V_z=30(1)\,E_{rz}$, where $E_{ri}=h^2/(2m\lambda_i^2)$, $i\in \{x,y,z\}$, are the corresponding recoil energies and $m$ is the mass of an atom. A magnetic field gradient $B'$ along $x$ is used to generate a linear potential of amplitude $\pm \Delta$ between neighboring sites, depending on the internal state of the atom [Fig.~\ref{Fig_Scheme}]. We use two Zeeman states with opposite magnetic moments denoted as spin-up $\ket{\uparrow}\equiv\ket{F=1,m_F=-1}$ and spin-down $\ket{\downarrow}\equiv\ket{F=2,m_F=-1}$. For $\Delta\gg J_x$, with $J_x$ being the bare coupling along $x$, tunneling is inhibited and can be restored resonantly using a pair of far-detuned running-wave beams with a frequency difference $\omega=\Delta/\hbar$ [Fig.~\ref{Fig_Scheme}]. The local optical potential created by these two beams is $V_K(\mathbf{r})=V_K^0 \textrm{cos}^2(\mathbf{q\,r}/2 + \omega t/2)$, with $\mathbf{q}=\mathbf{k_1}-\mathbf{k_2}$ being the wavevector difference. This gives rise to a time-dependent on-site modulation term with spatially-dependent phases $\phi_{m,n}=\mathbf{q\,R}$, where $\mathbf{R}=m \mathbf{d_x} + n \mathbf{d_y}$ denotes the lattice site $(m,n)$. In the high-frequency limit, $\hbar\omega\gg J_i$, $i\in\{x,y\}$, the system can be described by an effective time-independent Hamiltonian
\begin{eqnarray}
\hat H_{\uparrow,\downarrow}=&-&\sum_{m,n}\left(K\mathrm{e}^{\pm \ii \phi_{m,n}}\hat a^{\dagger}_{m+1,n}\hat a^{\phantom{\dagger}}_{m,n}
                     +J\hat a^{\dagger}_{m,n+1}\hat a^{\phantom{\dagger}}_{m,n}\right) \nonumber\\ 
                     &+&\mathrm{h.c.}\label{eq_Hamiltonian},
\end{eqnarray}
where the sign of the phase factor is positive for $\ket{\uparrow}$ atoms and negative for $\ket{\downarrow}$ atoms. In the limit of $\Delta\gg V_K^0$ the effective coupling strengths along $x$ and $y$ are $K=J_x \mathcal{J}_1\left(V_K^0/(\sqrt{2}\Delta)\right)\simeq J_xV_K^0/(2\sqrt{2}\Delta)$ and $J=J_y \mathcal{J}_0\left(V_K^0/(\sqrt{2}\Delta)\right)\simeq J_y$, where $\mathcal{J}_\nu(x)$ are the Bessel functions of the first kind~\cite{grossmann1992}. We note that the spin-dependent Peierls phase factors directly arise from the spin-dependent Zeeman coupling to the real applied magnetic field gradient.

For the chosen propagation of the running-wave beams shown in Fig.~\ref{Fig_Scheme} and $\lambda_K=2\lambda_y$, we obtain a phase factor $\phi_{m,n}=\pi/2 (m+n)$~\cite{supplements}. Therefore the phase accumulated on a closed path around a plaquette is $\pm\Phi=\pm\pi/2$, depending on the spin of the particle, and the corresponding gauge field is given by $\mathcal{A} = -(\hbar\Phi (x+y)/(d_x d_y) ,0,0) \, \hat \sigma_z$, where $\hat \sigma_z$ is the Pauli $z$-matrix. A different value of the flux $\Phi$ could be achieved by changing the wavelength $\lambda_K$ or the angle between the running-wave beams.

\begin{figure}[t!]
\includegraphics[width=\linewidth]{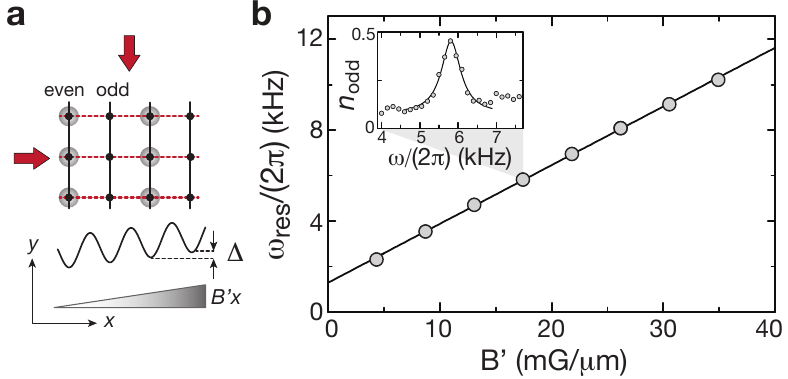}
\vspace{-0.cm} \caption{Laser-assisted tunneling in a tilted optical lattice.
\textbf{(a)} Schematic of the initial state used to study laser-assisted tunneling in the presence of a magnetic field gradient $B'$. Atoms are initially prepared in even sites with at most one atom per lattice site, while odd sites are left empty.  
\textbf{(b)} Measured frequency difference $\omega_{res}=\omega_1-\omega_2$, where atoms resonantly tunnel from even to odd sites, as a function of the magnetic field gradient $B'$. The finite value at $B'=0$ is due to a small additional magnetic field gradient \cite{supplements}. The solid line is a linear fit to our data. The inset shows a typical spectroscopy measurement for $B'=17.5\,\textrm{mG}/\mu\textrm{m}$, where the fraction of atoms on odd sites $n_{\textrm{odd}}$ is measured as a function of the frequency difference $\omega$ to determine the resonance frequency $\omega_{res}$. The solid line shows the fit of a Lorentzian-function to our data.
\label{Fig_LAT}} 
\end{figure}

To study laser-assisted tunneling in the presence of the magnetic field gradient $B'$, we loaded a Bose-Einstein condensate of about $5\times10^4$ atoms in an initial state, where all atoms populated even sites with at most one atom per site, while odd sites were left empty [Fig.~\ref{Fig_LAT}(a) and \cite{supplements}]. The final lattice depths, $V_x=5.0(1)\,E_{rx}$ and $V_y=40(1)\,E_{ry}$, were chosen to yield a negligible tunneling along $y$ and a bare tunnel coupling along $x$ of $J_x / h=0.26(1)\,\textrm{kHz}$. Due to the magnetic field gradient, tunneling was inhibited along $x$ and all atoms stayed in even sites. The running-wave beams were then switched on for 4~ms with strength $V_K^0=9.9(2)\,E_{rK}$, where $E_{rK}=h^2/(2m\lambda_K^2)$. Afterwards we measured the fraction of atoms transferred to odd sites $n_{\textrm{odd}}$ as a function of the frequency difference $\omega$ for a fixed value of the magnetic field gradient. Even-odd resolved detection was achieved by transferring atoms in odd sites to a higher Bloch band and applying a subsequent band-mapping sequence~\cite{supplements,sebbystrabley2006lattice}. As shown in the inset of Fig.~\ref{Fig_LAT}(b) atoms are transferred resonantly to odd sites when the frequency of the running-wave beams matches the energy offset $\Delta$ between neighboring sites. We measured the resonance frequency $\omega_{res}$ for various values of the magnetic field gradient and observed a large tunability up to about $\Delta/h\sim 10\,\textrm{kHz}$ [Fig.~\ref{Fig_LAT}(b)].

\begin{figure}[h]
\includegraphics[width=\linewidth]{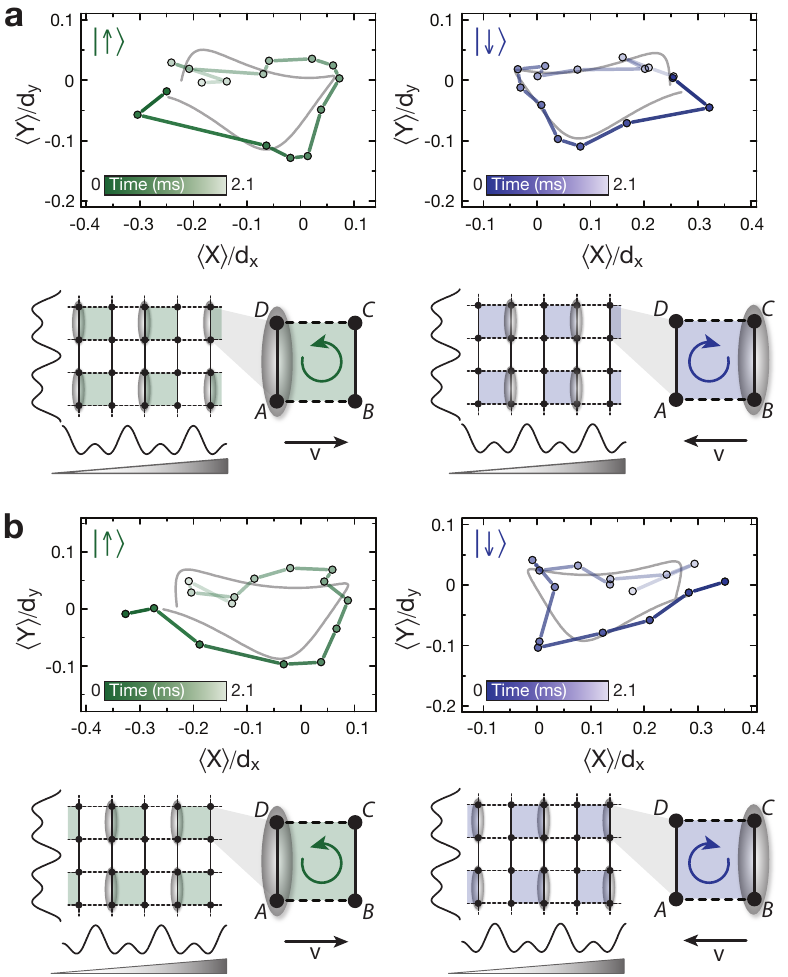}
\vspace{-0.cm} \caption{Quantum cyclotron orbits obtained from the mean atom positions along $x$ and $y$, $\left<X\right>/d_x$ and $\left<Y\right>/d_y$ for $J/K\approx2$ \cite{supplements}. Every data point is an average over three individual measurements. The solid gray lines show the fit of the theoretically expected evolution to the data, which was obtained from a numerical calculation solving the time-dependent Schr{\"o}dinger equation of the 4$\times$4 Hamiltonian. The oscillation amplitudes and offsets were fitted independently along $x$ and $y$, whereas the time offset $\tau=0.12(5)\,\textrm{ms}$ and flux $\Phi=0.73(5) \times \pi/2$ were fixed (see main text and \si{}~\cite{supplements}). The schematics illustrate the superlattice potentials used to partition the lattice into plaquettes together with the initial state for $\ket{\uparrow}$ atoms (green) and $\ket{\downarrow}$ atoms (blue) and the direction of the flux. The superlattice potential along $x$ is shifted by one lattice constant for the experimental results in \textbf{(b)} with respect to the ones in \textbf{(a)} to demonstrate the uniformity of the artificial magnetic field.\label{Fig_cyclotron}} 
\end{figure} 

The spatial distribution of the local fluxes induced by the running-wave beams was revealed by a series of measurements in isolated four-site square plaquettes using optical superlattices. This was achieved by superimposing two additional standing waves along $x$ and $y$ with wavelength $\lambda_{li}=2\lambda_i$, $i\in\{x,y\}$. The resulting potential along $x$ is $V(x)=V_{lx}\,\textrm{sin}^2(k_{x}x/2+\varphi_x/2)+V_{x}\,\textrm{sin}^2(k_{x}x)$, where  $V_{lx}$ is the depth of the ``long'' lattice. The superlattice potential along $y$ is given by an analogous expression.  The depths of the lattices and the relative phases, $\varphi_x$ and $\varphi_y$, can be controlled independently. For $\varphi_x=\varphi_y=0$ we realize symmetric double well potentials along $x$ and $y$ to isolate individual plaquettes [Fig.~\ref{Fig_cyclotron}]. Due to the presence of the magnetic field gradient, the plaquettes are tilted along $x$, with an energy offset $\Delta$ for $\ket{\uparrow}$ atoms and $-\Delta$ for $\ket{\downarrow}$ atoms. The four sites of the plaquette are denoted as $A, B, C, D$ [Fig.~\ref{Fig_cyclotron}]. The experiment started by loading spin-polarized single atoms into the ground state of the tilted plaquettes: $\ket{\Psi^0_{\uparrow}}=(\left|A\right>+\left|D\right>)/\sqrt{2}$ and $\ket{\Psi^0_{\downarrow}}=(\left|B\right>+\left|C\right>)/\sqrt{2}$, for $\ket{\uparrow}$ and $\ket{\downarrow}$, respectively [Fig.~\ref{Fig_cyclotron}(a) and \cite{supplements}]. After switching on the running-wave beams the atoms couple to the $B$ and $C$ sites ($\ket{\uparrow}$ atoms) and $A$ and $D$ sites ($\ket{\downarrow}$ atoms). Without the artificial magnetic field, the atoms would oscillate periodically between left and right, but due to the phase imprinted by the running-wave beams the atoms experience a force perpendicular to their velocity similar to the Lorentz force acting on a charged particle in a magnetic field. We measured the time evolution of the atom population on different bonds ($N_{\textrm{left}}=N_A+N_D$, $N_{\textrm{right}}=N_B+N_C$, $N_{\textrm{up}}=N_C+N_D$, and $N_{\textrm{down}}=N_A+N_B$), with $N_q$ being the atom population per site ($q=A, B, C, D$), by applying the even-odd resolved detection along both directions independently~\cite{supplements}. From this we obtained the mean atom positions along $x$ and $y$, $\left<X\right>=(N_{\textrm{right}}-N_{\textrm{left}})d_x/2N$ and $\left<Y\right>=(N_{\textrm{up}}-N_{\textrm{down}})d_y/2N$, with $N$ being the total atom number. As shown in Fig.~\ref{Fig_cyclotron}(a), the mean atom position follows a small-scale quantum analog of the classical cyclotron orbit for charged particles. Starting with equally populated sites $A$ and $D$, spin-up atoms experience a force along $y$, which is perpendicular to the initial velocity and points towards the lower bond in the plaquette ($A$ and $B$ sites). Spin-down atoms, initially with opposite velocity, also move towards the lower bond. Therefore the chirality of the cyclotron orbit is reversed, revealing the spin-dependent nature of the artificial magnetic field [Fig.~\ref{Fig_cyclotron}(a)]. The value of the magnetic flux per plaquette $\Phi=0.73(5)\times\pi/2$, measured in our previous work~\cite{aidelsburger2011experimental}, is used for the fits in Fig.~\ref{Fig_cyclotron}. The difference from $\Phi=\pi/2$, expected for a homogeneous lattice, stems from the smaller distance between lattice sites inside the plaquettes when separated.

To further demonstrate the uniformity of the magnetic field we performed the same set of measurements in plaquettes shifted by one lattice constant along $x$. This was achieved by changing the relative phase between the two standing waves along $x$ from $\varphi_x=0$ [Fig.~\ref{Fig_cyclotron}(a)] to $\varphi_x=\pi$ [Fig.~\ref{Fig_cyclotron}(b)]. The chirality of the obtained cyclotron orbits remained unchanged, which implies that a homogenous magnetic flux is present in the system.

\begin{figure}[t!]
\includegraphics[width=\linewidth]{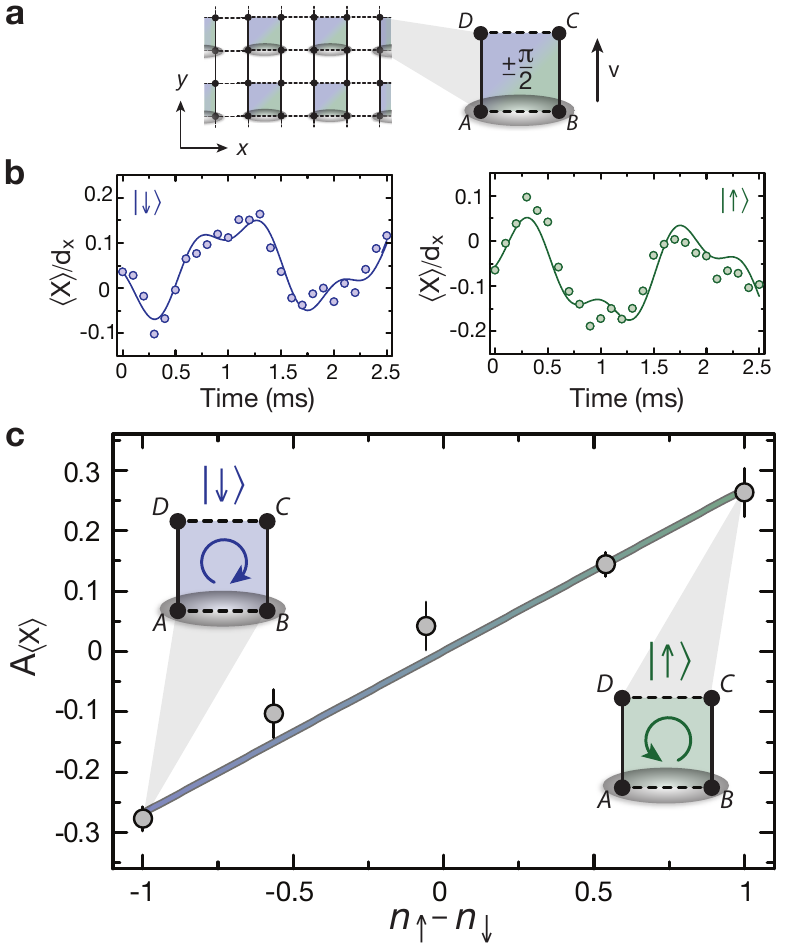}
\vspace{-0.cm} \caption{Particle currents as a function of the spin imbalance of the system.
\textbf{(a)} Schematic of the initial state configuration with one atom in a spin-mixture state of $\ket{\downarrow}$ and $\ket{\uparrow}$ in the lower bond of each plaquette \cite{supplements}.
\textbf{(b)} Evolution of the mean atom position along $x$ for all particles in $\ket{\downarrow}$ (blue) and all atoms in $\ket{\uparrow}$ (green) for $J/h=0.69(1)\,\textrm{kHz}$, $K/h=0.38(1)\,\textrm{kHz}$ and $\Delta/h=5.31(5)\,\textrm{kHz}$. The solid lines show a fit of the theoretically expected curve to the data obtained using the same method as in Fig.~\ref{Fig_cyclotron} with a time offset $\tau=0.18(3)\,\textrm{ms}$.
\textbf{(c)} Oscillation amplitude $A_{\left<X\right>}=A^{\downarrow}_{\left<X\right>}+A^{\uparrow}_{\left<X\right>}$ of the mean atom position $\left<X\right>/d_x$ as a function of the spin imbalance $n_{\uparrow}-n_{\downarrow}$. Every data point is an average over two individual measurements. The insets show schematics of the initial states together with the directions of the flux for all atoms in $\ket{\downarrow}$ (blue) and $\ket{\uparrow}$ (green). The solid line is a linear fit to our data, with the offset set to zero. The error bars show the standard deviation of our data.
\label{Fig_spin_mixture}} 
\end{figure}

In analogy to the spin Hall effect observed in solid state devices \cite{kato2004observation}, we measured the particle current perpendicular to the initial motion as a function of spin imbalance $n_{\uparrow}-n_{\downarrow}$, where $n_{\uparrow}$ ($n_{\downarrow}$) is the fraction of spin-up (-down) atoms. The experimental sequence started from a Mott insulator of unit filling, with each atom prepared in a superposition of $\ket{\uparrow}$ and $\ket{\downarrow}$. We then loaded single atoms into the ground-state of the lower bond of the plaquettes, which has an equal weight on $A$ and $B$ sites [Fig.~\ref{Fig_spin_mixture}(a)], and measured the mean atom position $\left<X\right>/d_x$. We obtained an oscillation amplitude for a spin-polarized state of $A_{\left<X\right>}= -0.28(2)$ for $\ket{\downarrow}$ atoms and $A_{\left<X\right>}= 0.26(4)$ for $\ket{\uparrow}$ atoms. As can be seen in Fig.~\ref{Fig_spin_mixture}(b), the evolution is almost perfectly mirrored for the two spin components. The measured oscillation amplitude as a function of the spin imbalance $n_{\uparrow}-n_{\downarrow}$ shows that the current depends linearly on the spin imbalance and reverses sign when flipping the spin [Fig.~\ref{Fig_spin_mixture}(c)]. 

In conclusion, we have demonstrated a new type of optical lattice that realizes the non-time reversal symmetric Hofstadter-Harper Hamiltonian and the time-reversal symmetric quantum spin Hall Hamiltonian for ultracold atoms in optical lattices. Loading spin-polarized or two-component Fermi gases into this lattice should allow one to directly realize quantum Hall and $Z_2$ topological insulators with chiral and helical edge states for finite sized systems. This system also opens the path to explore the fractal band structure of the Hofstadter butterfly with ultracold atoms~\cite{hofstadter1976energy}. The lowest band is topologically equivalent to the lowest Landau level and exhibits a Chern number of one~\cite{jaksch2003creation,mueller2004artificial,dalibard2010artificial,cooper2011optical}. In future experiments the ground-state properties, the effect of the Berry curvature, and topological edge states could be studied~\cite{baur2013adiabatic,price2012mapping,goldman2013direct}. The chiral edge modes in this lattice could be directly revealed in ladder systems exposed to a homogeneous magnetic field, which constitute the smallest possible 2D systems in which these states can be observed~\cite{huegel2013chiral}. Moreover, our work constitutes an important step towards the study of the quantum Hall effect with ultracold atomic gases and the creation of strongly-interacting fractional quantum Hall-like liquids for bosonic and fermionic atoms~\cite{sorensen2005fractional}.

We thank S.~Nascimb{\`e}ne and Y.-A.~Chen for their help in setting up the experiment and sharing of their ideas. This work was supported by the DFG (FOR635, FOR801), DARPA (OLE programm) and NIM. M. Aidelsburger was additionally supported by the Deutsche Telekom Stiftung. During the writing of the manuscript, we became aware of similar work carried out in the group of W.\,Ketterle at MIT~\cite{ketterle2013}.

\vspace{2cm}

\subsection*{Appendix}

\bigskip

\renewcommand{\thefigure}{A\arabic{figure}}
 \setcounter{figure}{0}
\renewcommand{\theequation}{A.\arabic{equation}}
 \setcounter{equation}{0}
 \renewcommand{\thesection}{A.\Roman{section}}
\setcounter{section}{0}

\section{Photon scattering rate}
A major advantage of our scheme is the large detuning of the running-wave beams from the atomic transition. For typical experimental parameters we estimated the photon scattering rate to be $2\times10^{-3}\,\textrm{s}^{-1}$ corresponding to a lifetime of the sample of several minutes, making the heating due to spontaneous emission negligible.

\section{Local phase distribution in the lattice}

\begin{figure}[thb]
\begin{center}
\includegraphics[scale=1]{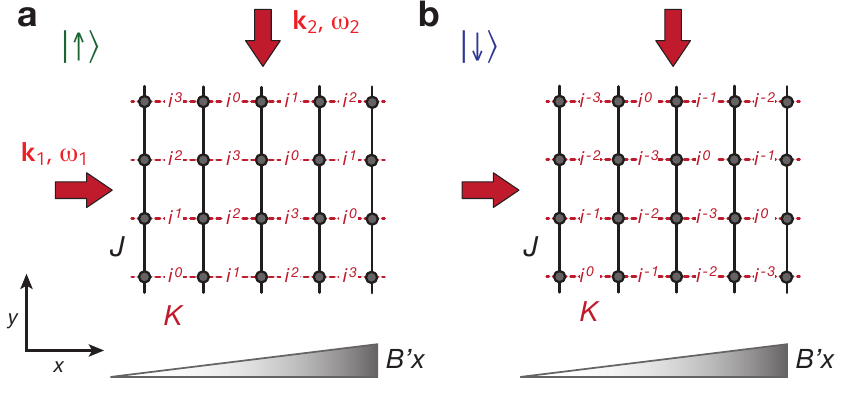}
\caption{Schematic drawing of the local phase distribution of the
  tunneling matrix elements induced by the pair of far-detuned
  running-wave beams used to restore resonant tunneling along $x$ for
  \textbf{(a)} $\ket{\uparrow}$ particles and \textbf{(b)}
  $\ket{\downarrow}$ particles.  }\label{fig:S1}
\end{center}
\end{figure}

As illustrated in Fig.~1 in the main text, bare tunneling with
amplitude $J$ occurs along the $y$ direction, while tunneling along
the $x$ direction is inhibited by a magnetic-field gradient $B'$. The
pair of far-detuned running-wave beams then restores tunneling along
that direction with complex amplitude $K$. The time-dependent optical
potential created by the interference between the two beams gives rise
to an on-site modulation with position-dependent phases
$\pm\phi_{m,n}=\pm \pi/2 (m+n)$, where the phase factors are positive for
$\ket{\uparrow}$ atoms and negative for $\ket{\downarrow}$ atoms. The
local distribution of phases in the lattice is illustrated in
Fig.~\ref{fig:S1}. For atoms in spin states $\ket{\uparrow}$
($\ket{\downarrow}$), the induced phase increases (decreases) linearly
along the vertical and horizontal directions resulting in a total flux
of $\Phi=\pi/2$ ($-\Phi$) per plaquette.

\section{Experimental sequence}

\subsection{Magnetic field gradient}
The magnetic field gradient $B'$ used to create the linear
potential is produced by a quadrupole magnetic field. Initially the
minimum of the quadrupole field is aligned with the position of the
atomic cloud. An additional single coil is then used to displace the
minimum of the magnetic trap along $x$ from the atom position by
generating a magnetic field offset $B_0$. However, the magnetic field
produced by the coil is not entirely homogeneous but causes a small
additional gradient $B'_0$. The resulting magnetic field gradient is
therefore given by the sum of the two: $B'+B'_0$. For the data
depicted in Fig.~2(b) of the main text we vary the strength of the
quadrupole field $B'$ without changing the offset field. The finite
value of the energy offset between neighboring sites of
$\Delta/h=1.30(4)\,\textrm{kHz}$ for $B'=0$ corresponds to $B'_0$.

\subsection{Initial state preparation on even sites}
The initial state for the laser assisted-tunneling measurements shown in Fig.~2 of the
main text, was prepared such that all atoms populated even sites
of the lattice with at most one atom per lattice site. We started by loading a Bose-Einstein condensate in the
$\left|F=1,m_F=-1\right>$-Zeeman state into a 3D optical lattice in
the Mott-insulator regime with at most two atoms per lattice site. The
lattice was created by the ``long lattice'' along $x$
($\lambda_{lx}=1534\,\textrm{nm}$), the ``short lattice'' along $y$
($\lambda_{y}=767\,\textrm{nm}$) and the vertical lattice with
wavelength $\lambda_{z}=844\,\textrm{nm}$. The lattice depths were
$V_{lx}=30(1)\,E_{rlx}$, $V_{y}=20.0(6)\,E_{ry}$ and $V_{z}=20.0(6)\,E_{rz}$, with $E_{ri}=h^2/(2m\lambda_i^2)$,
$i\in\{lx,y,z\}$, being the corresponding recoil energy. Afterwards we applied a filtering sequence where we
ramped up the lattices to $V_{lx}=104(3)\,E_{rlx}$,
$V_y=100(3)\,E_{ry}$, $V_z=120(4)\,E_{rz}$ and transferred all atoms
to the $\left|F=2,m_F=-1\right>$ state using a rapid adiabatic
transfer (Landau-Zener sweep). This was done by applying a microwave
field at about $6.8\,\textrm{GHz}$ together with a sweep of the
magnetic field offset over the resonance. The sweep was slow enough
($10\,\textrm{ms}$) to stay adiabatic and to achieve almost complete
population transfer. In the $\left|F=2,m_F=-1\right>$ state, spin
relaxation collisions are strongly enhanced. If two
atoms collide at least one of them is transferred to the $F=1$
manifold and enough energy is released to expel both atoms from the
trap \cite{myatt1997production,schmaljohann2004dynamics}. After
$50\,\textrm{ms}$ of waiting time nearly all atoms in doubly occupied
sites were removed from the trap and the remaining atoms were
transferred back to the $\left|F=1,m_F=-1\right>$ state using a
second rapid adiabatic transfer. Then we decreased the lattice depths
to $V_{lx}=52(2)\,E_{rlx}$, $V_y=40(1)\,E_{ry}$ and
$V_z=30(1)\,E_{rz}$, and ramped up the magnetic field gradient
within $200\,\textrm{ms}$. This time is chosen such that the magnetic
field gradient is stable over the duration of the experiment. After
that, the short lattice along $x$ was ramped up within
$20\,\textrm{ms}$ to $V_x=40(1)\,E_{rx}$ in order to load all atoms on
the lower energy wells of the tilted double well potentials. After
switching off the long lattice within $2\,\textrm{ms}$ we lowered the
short lattice to its final value of $V_x=5.0(1)\,E_{rx}$. This sequence
led to a state where all atoms only populated even sites with at most
one atom per lattice site.

\subsection{Sequence and parameters for the cyclotron orbits}
The experimental sequence for the cyclotron measurements shown in Fig.~3 of the main text started by loading a Bose-Einstein condensate in a 3D optical lattice in the Mott-insulator regime created by the two
long lattices along $x$ and $y$ and the vertical lattice at depths
$V_{i}=20.0(6)\,E_{ri}$, with $E_{ri}$ being the corresponding recoil energy and $i\in\{lx,ly,z\}$. For the filtering
sequence described above, we ramped up the lattices to
$V_{lx}=70(2)\,E_{rlx}$, $V_{ly}=70(2)\,E_{rly}$ and
$V_z=120(4)\,E_{rz}$. After all atoms in doubly occupied sites were lost from the trap we either performed a second Landau-Zener sweep to transfer all atoms to the $\left|\uparrow\right>=\left|F=1,m_F=-1\right>$ state (green data points) or we left all remaining atoms in the $\left|\downarrow\right>=\left|F=2,m_F=-1\right>$ state (blue data points).  Then we lowered the lattice depths to $V_{lx}=35(1)\,E_{rlx}$, $V_{ly}=35(1)\,E_{rly}$ and
$V_z=30(1)\,E_{rz}$ and subsequently ramped up the magnetic field gradient to its final value within $250\,\textrm{ms}$. After ramping up the short lattice along $x$ within $15\,\textrm{ms}$ to $V_x=40(1)\,E_{rx}$ all $\left|\uparrow\right>$ atoms were loaded in the left wells, while all $\left|\downarrow\right>$ atoms were loaded in the right wells. To load the ground-state in the tilted plaquettes we split the long lattice wells along $y$ by switching on the short lattice to $V_y=10.0(3)\,E_{ry}$ within $0.1\,\textrm{ms}$. At the same time we decreased the short lattice along $x$ to its final value $V_x=7.0(2)\,E_{rx}$. The dynamics were initiated by switching on the running-wave beams suddenly. The final trap parameters $J, K, \Delta$ were calibrated independently for each measurement. The corresponding values are given in the tables below:\\[3mm]

\noindent(a)\hspace{7mm}
\begin{tabular}{cccc}
spin & $J/h$ & $K/h$ & $\Delta/h$\\\hline
$\ket{\uparrow}$&$0.53(1)\,\textrm{kHz}$ & $0.27(1)\,\textrm{kHz}$ & $4.23(3)\,\textrm{kHz}$ \\
$\ket{\downarrow}$&$0.53(1)\,\textrm{kHz}$ & $0.28(1)\,\textrm{kHz}$ & $4.56(2)\,\textrm{kHz}$ \\
\end{tabular}\\[3mm]

\noindent(b)\hspace{7mm}
\begin{tabular}{cccc}
spin & $J/h$ & $K/h$ & $\Delta/h$\\\hline
$\ket{\uparrow}$&$0.53(1)\,\textrm{kHz}$ & $0.28(1)\,\textrm{kHz}$ & $4.91(4)\,\textrm{kHz}$ \\
$\ket{\downarrow}$&$0.53(1)\,\textrm{kHz}$ & $0.29(1)\,\textrm{kHz}$ & $4.36(2)\,\textrm{kHz}$ 
\end{tabular}\\[3mm]

\begin{figure}[th]
\begin{center}
\includegraphics[width=\columnwidth]{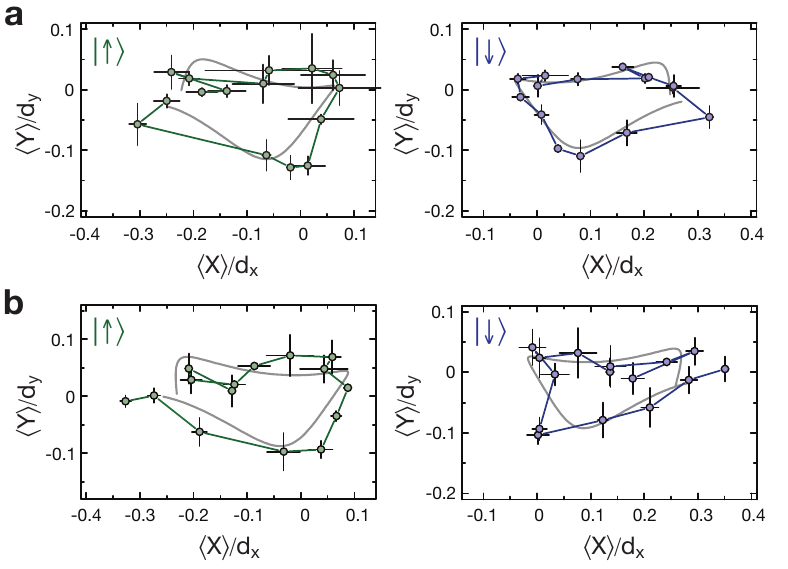}
\caption{Cyclotron orbits as shown in Fig.~3(a,b) of the main text. Each data point is averaged over three individual measurements. The error bars depict the corresponding standard deviation.}\label{fig:S2}
\end{center}
\end{figure}

The data points shown in Fig.~3 of the main text are averaged over three individual measurements. The corresponding standard deviations are depicted as error bars in Fig.~\ref{fig:S2}. Residual deviations of the data from the theory curve can most likely be attributed to an imperfect initial state preparation and off-resonant tunneling processes.

\subsection{Spin mixture and initial state preparation}
For the measurements shown in Fig.~4 of the main text, we loaded the atoms in a deep 3D optical lattice created by the two long lattices and the vertical lattice, $V_{lx}=30(1)\,E_{rlx}$,
$V_{ly}=20.0(6)\,E_{ry}$ and $V_z=20.0(6)\,E_{rz}$. For the filtering
sequence they were ramped up to $V_{lx}=104(3)\,E_{rlx}$, $V_{ly}=70(2)\,E_{rly}$ and
$V_z=120(4)\,E_{rz}$ and after all doubly occupied sites were lost from the trap, we were left with singly occupied sites and all atoms in the $\ket{\downarrow}$ state. In
contrast to the sequence described above, we stopped the second Landau-Zener sweep at different times to create a superposition of $\ket{\uparrow}$ and $\ket{\downarrow}$. The fraction of atoms in the 
$\ket{\uparrow}$ and $\ket{\downarrow}$ state was calibrated independently by applying a
Stern-Gerlach pulse at the beginning of the time of flight for each
sweep configuration. Afterwards we decreased the lattices to $V_{lx}=52(2)\,E_{rlx}$, $V_{ly}=35(2)\,E_{rly}$ and
$V_z=30(1)\,E_{rz}$. Similar to the sequence described above we ramped up the short lattice along $y$ within $20\,\textrm{ms}$ at a finite relative phase $\varphi_y\neq0$ in order to load the atoms in every second site of the superlattice potential along $y$. Tunneling along that direction was suppressed by the short lattice, $V_y=40(2)\,E_{ry}$. The relative phase was then adiabatically changed to $\varphi_y=0$ within $40\,\textrm{ms}$, to create a symmetric double well potential. After that we ramped up the magnetic field gradient within $250\,\textrm{ms}$ and subsequently switched on the short lattice along $x$ within $2\,\textrm{ms}$ to its final value $V_x=6.0(2)\,E_{rx}$. At this stage of the sequence $\ket{\uparrow}$ atoms were located in $A$ sites, while $\ket{\downarrow}$ atoms were located in $B$ sites. Tunneling along both directions was suppressed either by the magnetic field gradient ($x$) or a high potential barrier ($y$). Then the running-wave beams were switched on adiabatically within $7\,\textrm{ms}$ to load the atoms into the ground state of the lower bond of the plaquette, which has an equal weight on $A$ and $B$ sites. Afterwards the dynamics were initiated by lowering the barrier along $y$ within $100\,\mu\textrm{s}$ to its final value of $V_y=9.0(3)\,E_{ry}$.

\section{Site-resolved detection}
\begin{figure}[th]
\begin{center}
\includegraphics[width=\columnwidth]{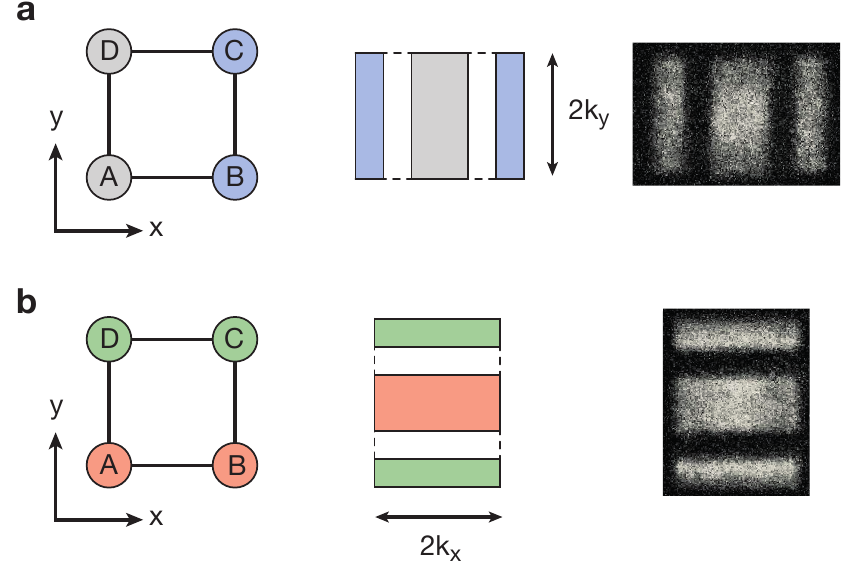}
\caption{Schematics of a four-site square plaquette and the
  corresponding 1-D long-lattice Brillouin zones after applying the
  mapping sequence \textbf{(a)} along $x$ and \textbf{(b)} along
  $y$. The grayscale images show a typical momentum distribution
  obtained after $10\,$ms of time-of-flight.  }\label{fig:S3}
\end{center}
\end{figure}

To detect the occupation numbers on the different bonds in the
plaquette, we apply two independent band mapping sequences along $x$ and $y$. In each experimental run only one of the two
sequences is applied. For the mapping along $x$, atoms in sites $B$
and $C$ are transferred to the third Bloch band of the long lattice
along x. A subsequent band mapping technique allows us to determine
the population in different Bloch bands by counting the corresponding
number of atoms
\cite{foelling2007direct,SebbyStrabley2007atomnumber}. The colors used
in Fig.~\ref{fig:S3}(a) show the connection between the long-lattice
Brillouin zones and the corresponding lattice sites for the mapping
along $x$, $N_{\textrm{left}}=N_A+N_D$ (gray) and
$N_{\textrm{right}}=N_B+N_C$ (blue). To obtain the population in the
lower and upper bonds we apply the same technique along
$y$. Figure~\ref{fig:S3}(b) illustrates the connection between the
lattice sites and Brillouin zones for the mapping sequence applied
along $y$, with $N_{\textrm{up}}=N_C+N_D$ (green) and
$N_{\textrm{down}}=N_A+N_B$ (red).

\section{Numerical simulations}

\subsection{Fitting functions of the time evolution of the mean atom positions}

\begin{figure}[th]
\begin{center}
\includegraphics[scale=1]{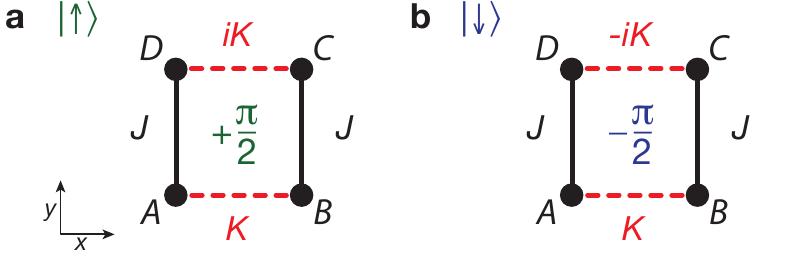}
\caption{Schematic drawings of a representative four-site square
  plaquette showing the tunnel couplings used in the $4\times4$
  Hamiltonian discussed in the text for \textbf{(a)} $\ket{\uparrow}$
  particles and \textbf{(b)} $\ket{\downarrow}$ particles.
}\label{fig:S4}
\end{center}
\end{figure}

In order to compare the measured evolutions of the mean atom positions
along $x$ and $y$ (shown in Fig.~3 and 4 of the main text) to the
theory, we simulate the dynamics of the system by solving the
time-dependent Schr{\"o}dinger equation numerically. The corresponding
Hamiltonian is a $4\times4$-matrix written in the single-particle
basis for each site $\{|A\rangle, |B\rangle, |C\rangle|, D\rangle\}$
with populations respectively given by $N_A,N_B,N_C,N_D$. The
parameters in this Hamiltonian are $J$, $K$ and $\Phi$. The tunnel
couplings $J$ and $K$ between neighboring sites are sketched in
Fig.~\ref{fig:S4} and their values were determined independently via
tunneling oscillations (see above). The flux was measured in our
earlier work~\cite{aidelsburger2011experimentalSM} and determined to be
$\Phi=0.73(5) \times \pi/2$. The obtained time evolutions $X(t)$ and
$Y(t)$ are then fitted to our experimental data using the following
functions $f_{X}(t)=X_0+A_{<X>}\cdot X(t+\tau)$ and
$f_{Y}(t)=Y_0+A_{<Y>}\cdot Y(t+\tau)$, with $X_0, A_{<X>}, Y_0,
A_{<Y>}$ being the fitting variables. A fixed time offset $\tau$ is
introduced, because the finite ramping times ($0.1-0.2\,\textrm{ms}$)
that initiate and freeze the dynamics prevent us from determining the
$t=0$ point a priori. To determine $\tau$, we fitted the evolution for
each data set independently and used the average values as fixed time
offsets for each figure.  For the data shown in Fig.~3 and 4, we
obtain average values of 0.12(5)~ms and 0.18(3)~ms, respectively,
which differ slightly because of different experimental sequences.

\subsection{Monte Carlo sampling of expected cyclotron orbits}

\begin{figure}[th]
\begin{center}
\includegraphics[width=83mm]{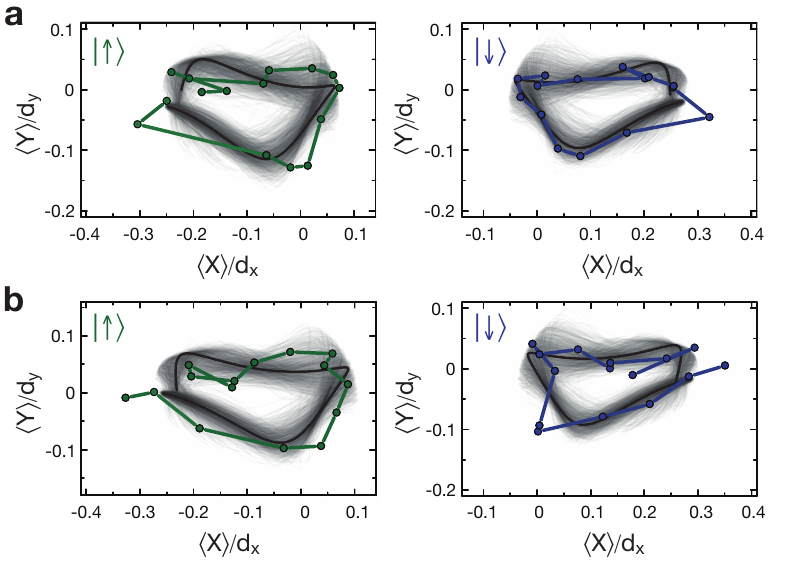}
\caption{Cyclotron orbits as shown and described in Fig.~3(a,b) of the
  main text, with experimentally measured cyclotron orbits (dots
  connected by green and blue solid lines), expected calculated
  evolution (black solid lines); but now each including 1000
  trajectories from Monte Carlo samples (gray lines) as decribed in
  the text.  }\label{fig:S5}
\end{center}
\end{figure}

Considering the uncertainty in our knowledge of the parameters $J$,
$K$, $\Phi$, and a potential detuning between lattice sites in each
direction of $\Delta_{x,y}=0(30)$ Hz, we can visualize the agreement
between the orbits derived from our model and those from our
measurements as follows. First, we assume each parameter $J$, $K$,
$\Phi$, and $\Delta_{x,y}$ follows a normal distribution with the mean
and standard deviation values reported above. Using these
distributions, we generate a sample of random parameter values that
are then used in our model Hamiltonian to calculate the cyclotron
orbits (a sampling procedure sometimes called Monte Carlo
sampling). The calculated orbit is then rescaled in amplitude, offset
and time delay as described above. By repeating this procedure with
1000 samples, we produce as many cyclotron orbits and plot them
together with the data in Fig.~\ref{fig:S5}.\\

\end{document}